\documentclass[prl,twocolumn,showpacs]{revtex4}

\usepackage{graphicx}
\usepackage{color}

\usepackage{amsmath,amssymb}
\usepackage{bm}
\usepackage{cancel}
\usepackage[normalem]{ulem}
\usepackage{upgreek}
\usepackage{hyperref}
\usepackage{float}

\newcommand{\PR}{Phys.~Rev.~}
\newcommand{\PRE}{Phys.~Rev.~E }
\newcommand{\PRL}{Phys.~Rev.~Lett.~}

\newcommand{\JPA}{J.~Phys.~A: Math.~Gen.~}

\newcommand{\EPJE}{Eur.~Phys.~J.~E }
\newcommand{\EPL}{Europhys.~Lett.~}
\newcommand{\JSTAT}{J.~Stat.~Mech.~(Theor.~Exp.) }

\newcommand{\st}{\text{s}}
\newcommand{\sig}{\widehat{\bm{\sigma}}}

\newcommand{\vv}{\bm{v}}
\newcommand{\hcs}{\text{HCS}}

\newcommand{\ini}{\text{ini}}

\begin{document} \title{Kovacs-like memory effect in driven granular
gases}
\author{A.\ Prados$^1,^2$ and E. Trizac$^2$}
\affiliation{$^1$
F\'{\i}sica Te\'{o}rica, Universidad de Sevilla, Apartado de Correos
1065, E-41080 Sevilla, Spain, EU}
\affiliation{$^2$ Universit\'e
Paris-Sud, Laboratoire de Physique Th\'eorique et Mod\`eles
Statistiques, UMR CNRS 8626, 91405 Orsay, France, EU}

\date{\today}
\begin{abstract}
  While memory effects have been reported for dense enough disordered
  systems such as glasses, we show here by a combination of analytical
  and simulation techniques that they are also intrinsic to the
  dynamics of dilute granular gases.  By means of a certain driving
  protocol, we prepare the gas in a state where the granular
  temperature $T$ coincides with its long time limit. However, $T$
  does not subsequently remain constant, but exhibits a non-monotonic
  evolution before reaching its non-equilibrium steady value.  The
  corresponding so-called Kovacs hump displays a normal behavior for
  weak dissipation (as observed in molecular systems), but is reversed
  under strong dissipation, where it thus becomes anomalous.
\end{abstract} 
\maketitle

At equilibrium, the response of a system to an external sudden
perturbation, like a temperature jump, depends only on the macroscopic
variables characterizing the state under study.  On the other hand,
in non-equilibrium situations, the observed response depends not only
on the instantaneous value of the macroscopic variables, but also on
the previous history. Memory effects are consequently ubiquitous out
of equilibrium. A classic experiment in this context bears the name of
Kovacs \cite{Ko63,Ko79}. A polymer sample, initially at equilibrium at
a high temperature $T_0$, is rapidly quenched to a low temperature
$T_1$, at which it evolves for a given \textit{waiting time}
$t_w$. Afterwards, the bath temperature is suddenly increased to $T$,
with $T_0>T>T_1$, such that the instantaneous polymer volume $V$
equals its equilibrium value at $T$. The sample volume then does not
remain constant for $t>t_w$: it first increases, displays a maximum,
and returns to equilibrium for longer times only.  This simple
experiment shows that the macroscopic variables $(P,V,T)$ (the
pressure $P$ being kept constant throughout the whole procedure) do
not completely characterize the macroscopic state of the system: its
response depends also on the previous \textit{thermal history}.

This kind of crossover, or Kovacs memory effect, has been extensively
investigated in glassy and other complex systems, starting from the
phenomenological theory presented by Kovacs himself \cite{Ko79}. It is
displayed by polymers, structural and spin glasses, compacting dense
granular media, kinetically constrained models, classical and quantum
spin models, distributions of two-level systems,
etc.~\cite{Ko63,Ko79,Br78,ByB02,ByH02,Bu03,BBDG03,CLyL04,AyS04,MyS04,TyV04,ALyN06,AAyN08,PyB10,ByL10,DyH11,RyP14}. The
quantity displaying the hump may be different from the volume: in
several of the previous studies, the energy is the relevant
quantity. Interestingly, most of the observed behavior can be
understood within a linear response theory approach, although the
temperature jumps are usually not small in the experiments
\cite{PyB10,DyH11,RyP14}.

Whereas the Kovacs effect has previously been reported for dense
media, or systems exhibiting complex energy landscape, we focus here
on a low density granular gas \cite{PyB03,ByP04} where the effect is
{\it a priori} less expected.  Due to inelastic collisions, a gas of
grains is an intrinsically out-of-equilibrium system, arguably one of
the simplest.  Without external driving, its granular temperature -- a
measure of velocity fluctuations -- monotonically decreases, and the
granular gas may end up in the homogeneous cooling state (HCS),
provided a small enough system is considered to prevent the
development of long-wavelength instabilities
\cite{GyS95,BRyC96,NyE98}. In order to reach a non-equilibrium steady
state, one needs a mechanism that inputs energy into the set-up. With
the \textit{stochastic thermostat} \cite{NyE98,ENTyP99}, additional
white noise forces act over each grain independently. This simple
forcing mechanism is relevant for some two-dimensional experimental
configurations with a rough vibrating piston \cite{PEyU02}, and also
appears as a limiting case of a granular system heated by elastic
collisions \cite{Sa03}. Although these thermostatted or heated
granular fluids have been extensively investigated
\cite{WM96,NyE98,ENTyP99,MyS00,SyM09,MGyT09,ETB06,Z09,GMyT12,GMyT13,P98,CVyG13},
no attention has been paid to memory effects. On the other hand, in
compaction processes of dense granular systems, the relevance of
history has been assessed, both experimentally and theoretically: Its
evolution under a given driving depends not only on the instantaneous
value of its packing fraction, but also on the previous driving
protocol
\cite{TyV04,JTMyJ00,He00,ByP01,ByL01,ByP02,RDRNyB05,RRPByD07}.

A valid question in granular gases is the type and number of variables
that completely characterize a macroscopic state \cite{rque20}. In the
non-driven case, the HCS is the reference state for developing the
hydrodynamics, and it suffices to give the granular temperature. The
same holds for the Gaussian thermostatted case
\cite{MyS00,Lu01,BRyM04}, which can be mapped onto the HCS. On the
other hand, there is some evidence that additional variables are
necessary for other drivings like the stochastic thermostat. This
uniformly heated granular gas evolves to a hydrodynamic
solution of the Boltzmann equation \cite{GMyT12,GMyT13}, the so-called
$\beta$-state where $\beta$ is a parameter that keeps track of the
distance to stationarity (see below). Therein, the granular
temperature is a monotonic function of time and, together with the
driving intensity, completely characterizes the $\beta$-state. One may
thus naively conclude that no Kovacs hump should be expected. We show
below that such a surmise is incorrect: not only is the Kovacs effect
present, but it also changes sign depending on dissipation.  An
anomalous Kovacs effect is thereby brought to bear for strongly
dissipative systems.

In short, our motivation is two-fold. First, adapting the celebrated
Kovacs protocol, we wish to study if memory can be encoded in a
seemingly plain system with a trivial energy landscape, which is all
kinetic.  Second, the goal is to illustrate for the fact that, for a
given driving amplitude, a single index (temperature) is insufficient
to describe the non-equilibrium behavior of our homogeneous gas.  One
must keep track also of the non-Gaussianities of the velocity
fluctuations, through the excess kurtosis.  It appears that these
non-Gaussianities are necessary, although not sufficient in general,
for the occurrence of the hump.

\begin{figure}
\centering 
  \includegraphics[width=3.in]{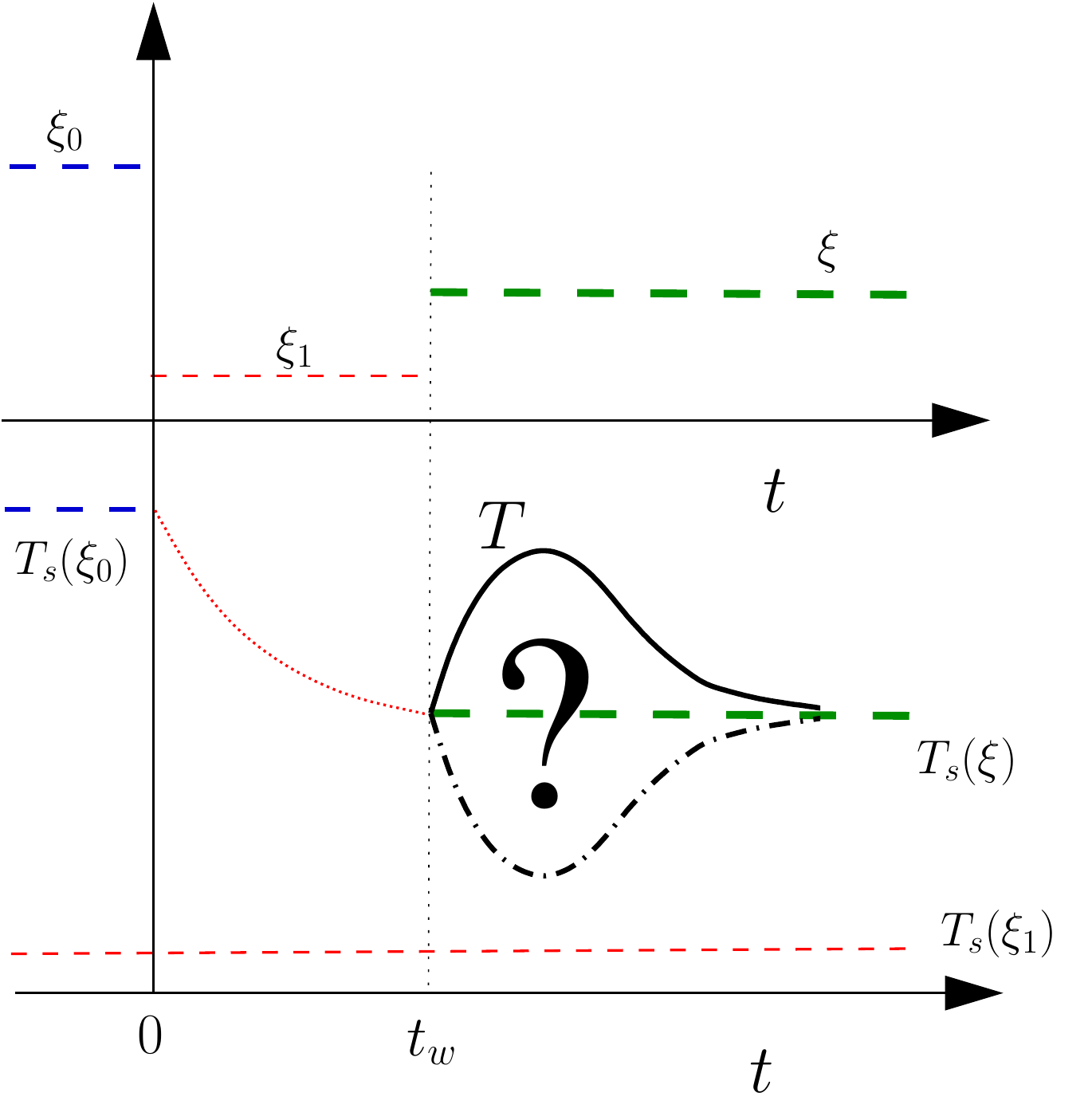}\\
  \caption{Top: Sketch of the drive time dependence.  Bottom: Ensuing
    temperature evolution.  At $t=0$, the gas is at temperature
    $T_s(\xi_0)$, in the non-equilibrium steady state corresponding to
    a value of the driving $\xi_0$. At $t=0$, the drive is
    suddenly decreased to $\xi_1\ll\xi_0$, which is kept for a waiting
    time $t_w$. At $t=t_w$, the granular temperature is
    measured, and the driving is cranked up to a new value $\xi$, such
    that $T_\st(\xi)=T(t_w)$. The question mark is for the two
    possible scenarios: a positive hump with a $T$ maximum (normal
    behavior, solid line), or a negative anomalous hump (dot-dashed
    line). At long times, $T$ reaches its steady value
    $T_s(\xi)$. }\label{fig:protocol}
\end{figure}

The system at hand comprises $N$ inelastic smooth hard particles of
mass $m$ and diameter $\sigma$. When particles $i$ and $j$ collide,
momentum is conserved but kinetic energy is not.  The inelasticity is
characterized by the coefficient of normal restitution $\alpha$ (taken
independent of the relative velocity):
$\bm{\sigma}\cdot\vv'_{ij}=-\alpha\,\bm{\sigma}\cdot\vv_{ij}$, in which
$\vv'_{ij}$ is the post-collisional relative velocity, $\vv_{ij}$ the
pre-collisional one, and $\sig$ the unit vector joining the centers of
particles $j$ and $i$.  Moreover, grains are submitted to independent
white noise forces, and we assume that the system remains spatially
homogeneous, as backed up by molecular dynamics simulations
\cite{ENTyP99}. Then, the velocity probability distribution is a sole
function of velocity and time, and obeys \cite{NyE98,ENTyP99,Sa03},
\begin{eqnarray}
 \partial_t f(\vv_1,t)=\sigma^{d-1}\!\int d\vv_2 \int d\sig\,
\Theta(\vv_{12}\cdot\sig)
(\vv_{12}\cdot\sig) &&
 \nonumber \\
 \qquad\times(\alpha^{-2}b_{\sigma}^{-1}-1)
f(\vv_1,t)f(\vv_2,t)+\frac{\xi^2}{2}\nabla^2_{\vv_1} f(\vv_1,t)&&
\label{1.2}
\end{eqnarray}
In the Boltzmann-Fokker-Planck
equation above, $\xi$ is the noise strength, $d$ is the
dimension of space, $\Theta$ is Heaviside function, and the operator
$b_{\sigma}^{-1}$ replaces the velocities $\vv_1$ and $\vv_2$ by the
pre-collisional ones.

The granular temperature $T(t)$ is defined as the second moment of the
distribution,
\begin{equation}\label{1.4} n \left\langle\frac{1}{2}m
v^2(t)\right\rangle \equiv \int d\vv \frac{1}{2}mv^2
f(\vv,t)=\frac{d}{2} n T(t),
\end{equation} where $n=\int d\vv f(\vv,t)$ is the particle density. In the theory
developed here, a central role is played by the \textit{excess
kurtosis} $a_2$ of the velocity fluctuations,
\begin{equation}\label{1.5} a_2=\frac{d}{d+2} \frac{\langle
v^4\rangle}{\langle v^2\rangle^2}-1,
\end{equation} which vanishes for a Gaussian distribution. The general
$n$-th moment is given by $ \langle v^n\rangle \equiv n^{-1}\int d\vv
\, v^n f(\vv,t)$. In the long time limit,
the granular gas reaches a steady state in which the energy loss due
to collisions is balanced on average by the energy input from the
stochastic thermostat. The stationary values of the granular
temperature $T_\st$ and excess kurtosis $a_2^\st$ are \cite{NyE98}
\begin{subequations}
\begin{equation}\label{1.11} T_\st=\left[\frac{m\xi^2}
{\zeta_0(1+\frac{3}{16}a_2^\st)}\right]^{2/3}, \quad \zeta_0=\frac{2 n
\sigma^{d-1} \left(1-\alpha^2\right) \pi^{\frac{d-1}{2}}}{\sqrt{m}
d\Gamma(d/2)},
\end{equation}
\begin{equation}\label{1.10} a_2^\st=\frac{16(1-\alpha)(1-2\alpha^2)}
{73+56d-24d\alpha-105\alpha+30(1-\alpha)\alpha^2}.
\end{equation}
\end{subequations}
The main assumptions in deriving these steady values are (i) the
first-Sonine approximation (ii) the smallness of non-linear terms in
the excess kurtosis, which are thus neglected (see e.g. \cite{NyE98}).
For our purposes, it is convenient to introduce rescaled,
order of unity variables,
\begin{equation}\label{1.12}
\beta=\sqrt{\frac{T_\st}{T}}, \quad
A_2=\frac{a_2}{a_2^\st}, \quad \uptau=\frac{\zeta_0 \sqrt{T_\st}}{2}
t.
 \end{equation}
Starting from the Boltzmann-Fokker-Planck equation
\eqref{1.2}, one can derive the evolution equations for the granular temperature and the excess kurtosis
\cite{NyE98,GMyT12,PyT13},
\begin{subequations}\label{1.17}
\begin{equation}\label{1.17a} \frac{d\beta}{d\uptau}
= 1-\beta^3
+\frac{3}{16}a_2^\st \left( A_2-\beta^3 \right),
\end{equation}
\begin{equation}\label{1.17b} \beta \frac{dA_2}{d\uptau}
= 4 \left[
\left(1-\beta^3\right) A_2+ B \left( 1-A_2 \right) \right],
\end{equation}
\end{subequations} which are nonlinear in $\beta$ but linear in the
excess kurtosis, consistently with our approach. Obviously, $\beta=1$
and $A_2=1$ is a stationary solution. The parameter $B$ is a given
function of the restitution coefficient and of the dimension of space.
We find it from a self-consistency argument: when
the driving is so small that $\beta\to 0$, $a_2$
evolves to its value $a_2^\hcs$  for the HCS \cite{SyM09},
\begin{equation}\label{2.3}
a_2^\hcs=\frac{16(1-\alpha)(1-2\alpha^2)}{25 +2\alpha(\alpha-1) + 24 d + \alpha ( 8 d-57 )}.
\end{equation}
Thus, $A_2=a_2^\hcs/a_2^\st$ should be a root of the right hand side
of Eq. (\ref{1.17b}), and $B=a_{2}^{\hcs}/(a_{2}^{\hcs}-a_{2}^{\st})$,
that is,
\begin{equation}\label{1.15}
B \,=\,\frac{73+8d(7-3\alpha)+15\alpha[2\alpha(1-\alpha)-7]}
{16(1-\alpha)(3+2d+2\alpha^2)}.
\end{equation}

Let us address the Kovacs-like experiment depicted in
Fig.~\ref{fig:protocol}. We would like to investigate the behavior of
the granular temperature $T$ for $t>t_w$. If the pair $(\xi,T)$ does
not completely characterize the state of the system, and other
variables should be taken into account, $T$ will not remain constant
but separate from its steady (initial) value and have either a maximum
or a minimum. In molecular systems, there always appears a maximum in
the Kovacs hump.  This does not have to be the case for the granular
temperature, because the granular gas is an intrinsically dissipative,
out-of-equilibrium, system.

\begin{figure}
 \centering 
  \includegraphics[width=3.25in]{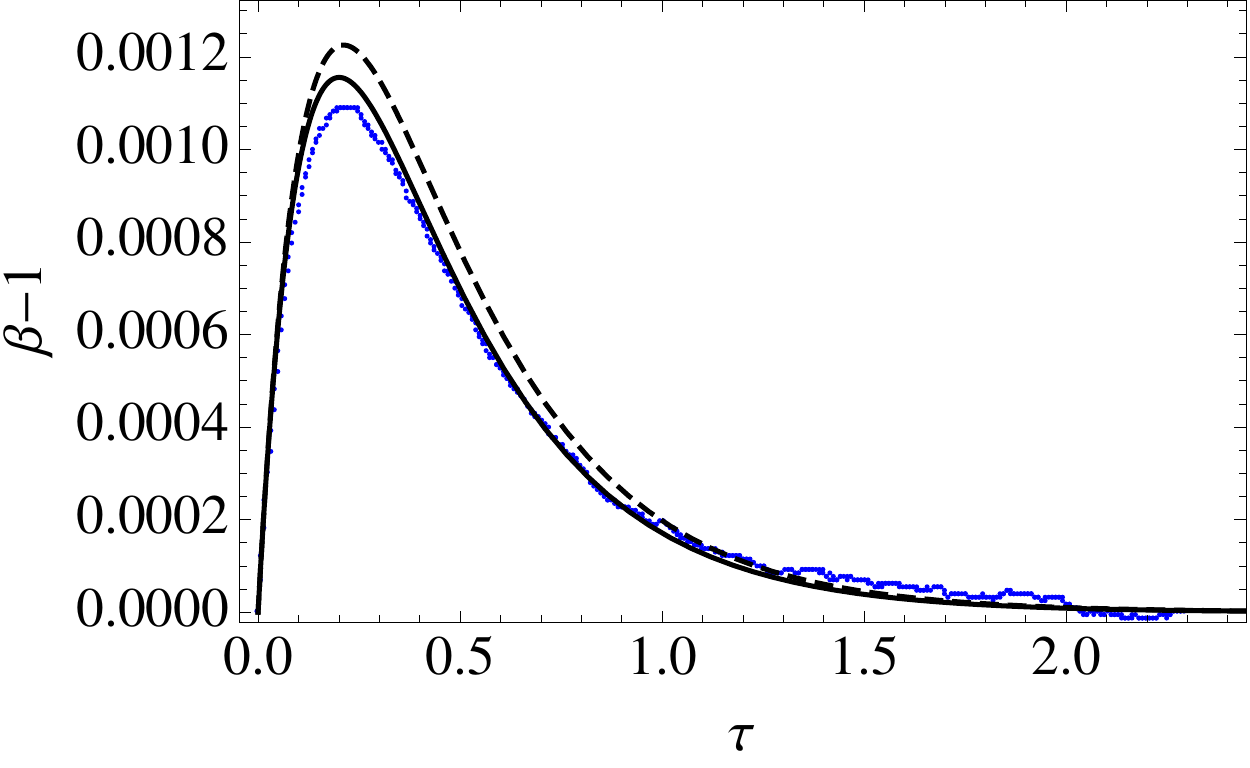}\\
  \includegraphics[width=3.25in]{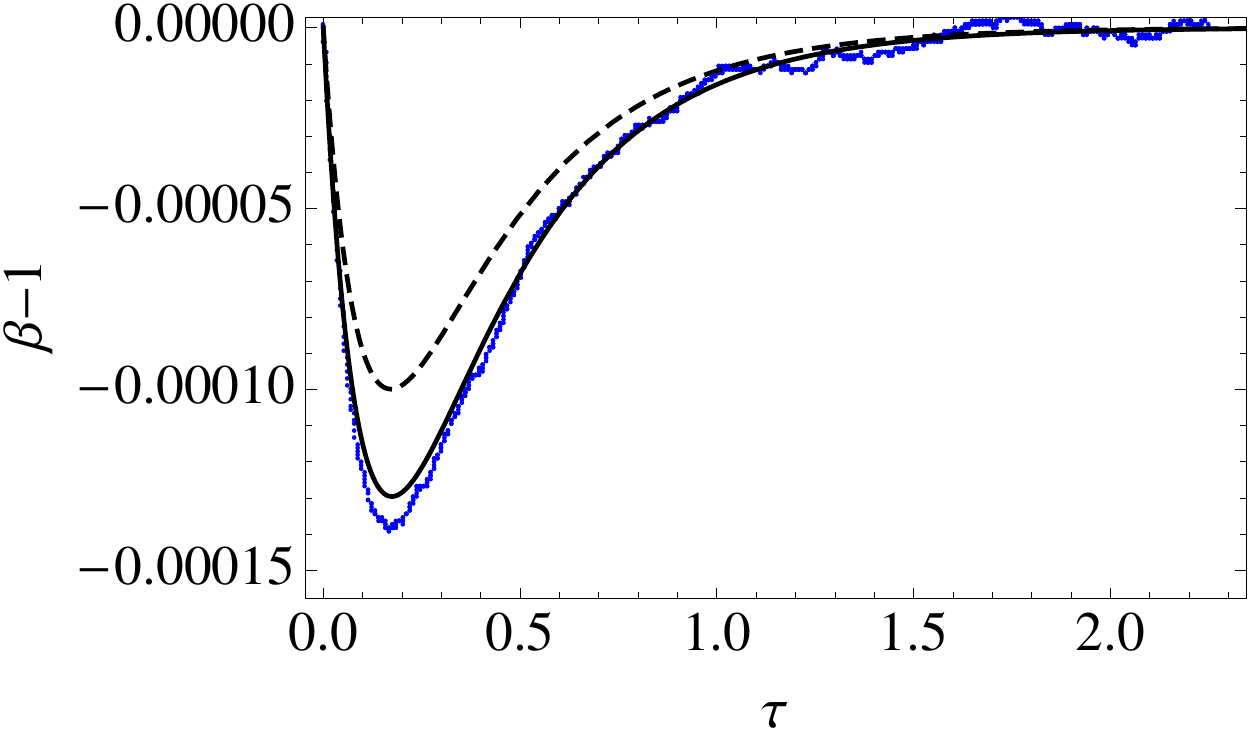}\\
  \caption{
    Plot of the Kovacs hump for $\alpha=0.3$ (top) and
    $\alpha=0.8$ (bottom). Monte Carlo simulation curves (points) for
    a system of $10^{4}$ hard disks ($d=2$),
    averaged over $10^5$ (top) and $1.5\times 10^6$
    trajectories (bottom).  They are compared to the theoretical curve
    \eqref{2.13}: the dashed line corresponds to the predicted values
    of $a_2^\st$, $a_2^\ini$ and $B$, while the solid line is obtained
    by taking these three parameters from the simulation (see e.g.
    Fig. \ref{fig:a2decay}, from which $B$ is directly measured).  The
    sign of $\beta-1$ changes from the highly inelastic (top) to the
    weakly inelastic (bottom) case. Note that a maximum of $\beta$
    corresponds to a minimum of $T=T_\st/\beta^2$ (and vice versa), so
    that the Kovacs hump is anomalous in the highly inelastic region.
  }\label{fig2}
\end{figure}

Defining the shifted time variable
$\uptau=\zeta_0\sqrt{T_s}(t-t_w)/2$, we have to solve
Eqs.~\eqref{1.17} with the initial conditions $\beta(\uptau=0)=1$ and
$A_2(\uptau=0)=a_2^\ini/a_2^\st$, where $a_2^\ini$ is the value of the
excess kurtosis in the final state of the waiting time window. Since
$a_2^\st$ is small ($|a_2^\st|\leq 0.07$) across the whole range of
restitution coefficients, while $\beta$ and $A_2$ are of the order of
unity, we expand both $\beta$ and $A_2$ in powers of $a_2^\st$ to
obtain an approximate solution of Eqs.~\eqref{1.17}\cite{PyT13},
\begin{subequations}\label{result}
\begin{equation}\label{2.8b} \quad a_{2}(\uptau)-a_2^\st \, \sim \,
(a_2^\ini-a_2^\st) \, e^{-4B\uptau},
\end{equation}
\begin{equation}\label{2.13} \beta(\uptau)-1 \, \sim \, \frac{3 \left(a_2^\ini-a_2^\st \right)}{16(4B-3)}
 \left( e^{-3\uptau}-e^{-4B\uptau}
\right).
\end{equation}
\end{subequations}
The relaxation of the excess kurtosis to its steady value is
exponential, while that of the rescaled temperature $\beta$ is the sum
of two exponentials with different relaxation times.  The sign of
$\beta-1$ is the same as that of $a_{2}^{\st}$ because (i) $4B>3$ and
(ii) $(a_2^\ini-a_2^\st)$ and $a_2^\st$ have the same sign as a
function of the restitution coefficient for the arbitrary ``cooling''
($\xi_{0}>\xi>\xi_{1}$) protocol in Fig.~\ref{fig:protocol}.  In fact,
Eq.~\eqref{1.17b} predicts that $dA_{2}/d\uptau$ is initially positive
and thus $|a_{2}|>|a_{2}^{\st}|$ in the whole waiting time window
\cite{PyT13}.  In addition, the steady excess kurtosis $a_{2}^{\st}$
changes sign at $\alpha_c = 1/\sqrt{2}\simeq 0.707$: $a_2^\st>0$ for
$\alpha<\alpha_c$ while $a_2^\st<0$ for $\alpha>\alpha_c$
\cite{alpha_c}. Thus, for small inelasticity ($\alpha>\alpha_c$),
$\beta-1<0$ and $\beta$ has a minimum, while the granular temperature
$T=T_\st/\beta^2$ has a maximum. This behavior is completely similar
to that of glassy systems, so we may speak of a \textit{normal} Kovacs
hump in the weakly dissipative case. On the contrary, for high
inelasticity, $\alpha<\alpha_c$, $\beta-1>0$ and $\beta$ displays a
maximum, which corresponds to a minimum of $T$: an \textit{anomalous}
Kovacs hump appears.

\begin{figure}
 \centering 
  \includegraphics[width=3.25in]{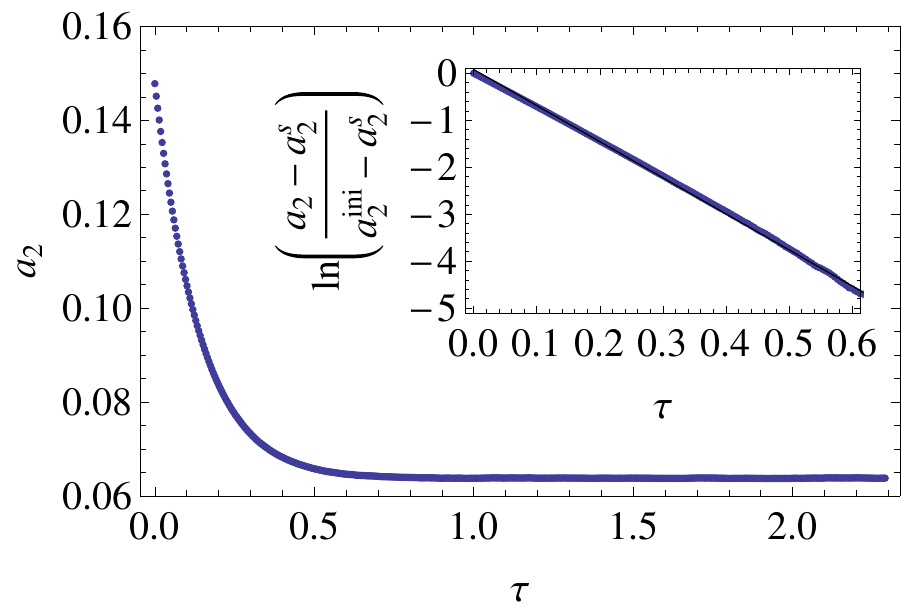}\\
  \caption{
    Decay of the excess kurtosis from its initial to its
    steady state value $a_2^\st$. Plotted is the simulation curve
    obtained by the direct Monte Carlo scheme for $\alpha=0.3$.  In
    the inset, the same decay but on a logarithmic scale. The linear
    slope is directly related to the parameter $B$, see \eqref{2.8b}.
  }\label{fig:a2decay}
\end{figure}

In Fig. \ref{fig2}, the above theoretical prediction for the Kovacs
hump is tested against numerical computations.  The latter are
obtained by means of direct Monte Carlo simulations \cite{Bi94} of the
Boltzmann-Fokker-Planck equation (\ref{1.2}).  Two values of the
restitution coefficient are considered: (i) $\alpha=0.3<\alpha_c$
(top, high inelasticity), and (ii) $\alpha=0.8>\alpha_c$ (bottom, low
inelasticity). For the sake of concreteness, we take the limiting case
(i) $\xi_1=0$ (the granular gas freely cools in the time window
$0<t<t_w$) and (ii) a long enough $t_w$, so that
$a_{2}^{\ini}=a_{2}^{\hcs}$. This choice (i)-(ii) is somewhat
immaterial for what follows, because the whole dependence of the
Kovacs hump on $\xi_{1}$ and $t_{w}$ is encoded in the initial value
of the excess kurtosis difference $a_{2}^{\ini}-a_{2}^{\st}$, which in
turn only changes the scale of the hump but does not alter its shape
\cite{rque50}. In both cases, the dashed line corresponds to the
theoretical prediction, Eq.~\eqref{2.13}, in which the values of
$a_2^\st$, $B$, and $a_2^\hcs$ are given by Eqs. \eqref{1.10},
\eqref{2.3} and \eqref{1.15}, respectively. The agreement is
reasonable: in particular, the sign of the hump is correctly
predicted, but there are quantitative discrepancies. The latter stem
from errors (of up to $10\%$) in the theoretical estimates of $a_{2}$
and $B$ \cite{GMyT12}. The quantitative agreement can be improved by
inserting into \eqref{2.13} their simulation values \cite{PyT13},
which yields the solid line.  In particular, $B$ is extracted from
Fig. \ref{fig:a2decay}, which furthermore corroborates the prediction
of Eq. (\ref{2.8b}).

In order to understand the physical mechanism responsible for the
observed behavior, a central idea is that the energy dissipation rate
$d$ (``cooling rate'' in the granular gas literature) increases with
the excess kurtosis \cite{NyE98}. Moreover, the unforced system has
stronger non-Gaussanities than the driven one,
$|a_{2}^{\hcs}|>|a_{2}^{\st}|$ \cite{additional_friction}, because the
latter is randomized from stochastic 'kicks' due to the forcing. For
small inelasticities ($\alpha>\alpha_{c}$), $a_{2}^{\hcs}$ and
$a_{2}^{\st}$ are both negative, so that $a_{2}^{\hcs}<a_{2}^{\st}$
and at $t=t_{w}$ the system has the steady value of the granular
temperature but a dissipation rate smaller than that at stationarity
$d_{\st}$, $d/d_{\st}<1$. Therefore, the granular temperature $T$
first increases and passes through a maximum ($\beta$ minimum) before
returning to its steady value. For high inelasticities
($\alpha<\alpha_{c}$), $a_{2}^{\hcs}$ and $a_{2}^{\st}$ are both
positive, so that $a_{2}^{\hcs}>a_{2}^{\st}$. Then, the system is at
$t=t_w$ transiently in a state with $d/d_{\st}>1$, so that $T$
initially decreases and passes through a minimum ($\beta$ maximum),
see the table.

\begin{table}
  \centering
  \begin{tabular}{|c|c|c|c|c|c|}
\hline
    inelasticity & $\alpha$ & $a_{2}^{\hcs}-a_{2}^{\st}$ & $d/d_{\st}$ &
    $T$ hump (Kovacs) \\
\hline
\hline
    ``low''& $\; >\alpha_{c} \;$ & $\; <0 \;$ & $\; <1 \;$ &
    maximum (normal) \\
``high''& $\; <\alpha_{c} \;$ & $\; >0 \;$ & $\; >1 \;$ &
    minimum (anomalous) \\
\hline
  \end{tabular}
  \caption{\label{table} Summary of the Kovacs hump phenomenology and
    the underlying physical mechanism for the driving protocol in
    Fig.~\ref{fig:protocol},
    with $\xi_{1}\ll \xi_0$.}
\end{table}

The existence of the Kovacs hump, as given by Eq.~\eqref{2.13}, is a
crisp proof that the granular temperature does not suffice for
characterizing the state of uniformly heated granular gases.
Moreover, it links granular gases and other complex, non-equilibrium,
systems.  Nevertheless, this crossover effect is not a direct
extension of the similar phenomenon observed in the latter: here we
are dealing with an intrinsically out of equilibrium system relaxing
to a far from equilibrium steady state. Furthermore, for the protocol
considered, the intrinsically dissipative dynamics makes the Kovacs
hump anomalous for high inelasticity.  The hump is normal for the
weakly dissipative case and disappears in the elastic limit $\alpha\to
1$, in which both $a_2^\hcs$ and $a_2^\st$ vanish. If we considered a
``heating'' protocol, that is, $\xi_{0}<\xi<\xi_{1}$,
Eq.~\eqref{1.17b} would give that $dA_{2}/d\uptau$ is initially
negative: $|a_{2}|<|a_{2}^{\st}|$ in the waiting time window. Then,
$a_{2}^{\ini}-a_{2}^{\st}$ would have the sign opposite to that of
$a_{2}^{\st}$ and the sign of the hump would be reversed as compared
to the behavior shown in the table. Here again, the normal behavior
appears for low inelasticity, since in molecular systems, the energy
displays a minimum for such ``heating'' protocols \cite{DyH11}.

Provided that the first-Sonine approximation to the Boltzmann equation
remains valid, some of our main results are expected to hold for
almost any uniformly heated granular gas: (i) the proportionality of
the hump to the difference of excess kurtosis $(a_2^\ini-a_2^\st)$,
(ii) the exponential relaxation of the excess kurtosis, (iii) the
two-exponential structure of the granular temperature relaxation.  A
singular case would be that of the Gaussian-thermostatted system,
which can be mapped onto the HCS: In particular, its excess kurtosis
equals $a_2^\hcs$ and no hump would be observed. This is consistent,
since the granular temperature completely specifies the HCS. Moreover,
this clearly shows that the generic non-Maxwellian ($a_2\neq 0$)
character of the velocity distribution function of granular gases is
not a sufficient condition for the existence of the crossover effect.

The formalism developed here is thus quite general and may open the
door to further general results in non-equilibrium statistical
physics. In particular, the anomalous Kovacs hump for high
inelasticity deserves further investigation. Linear response results
\cite{PyB10,DyH11,RyP14}, closely related to the fluctuation-dissipation
theorem, assure that the Kovacs hump is normal in molecular
systems. In this regard, it would be interesting to analyze the
possible connection between this anomaly and the validity of
fluctuation-dissipation-like relations in dissipative systems
\cite{PByL02,PByV07,MGyT09,PLyH11-12,BMyG12}.

\acknowledgments

We acknowledge useful discussions with M.I. Garc\'{\i}a de Soria and
P. Maynar. This work has been supported by the Spanish Ministerio de
Econom\'\i a y Competitividad grant FIS2011-24460 (AP). AP would also
like to thank the Spanish Ministerio de Educaci\'on, Cultura y Deporte
mobility grant PRX12/00362 that funded his stay at the Universit\'e
Paris-Sud in the summer of 2013, during which this work was carried
out.


\begin{thebibliography}{99}
\bibitem{rque20} Note that we will address here homogeneous configurations, without
 macroscopic flow.
\bibitem{alpha_c} The accuracy of this theoretical value of
  $\alpha_{c}$ has been checked numerically, see for instance
  \cite{MyS00}.
\bibitem{rque50} For a finite value of $\xi_1/\xi_0$, there exists an
  optimal waiting time $t_w$ for which $|a_{2}^{\ini}-a_{2}^{\hcs}|$
  is maximal. For $\xi_{1}\ll \xi_{0}$, the optimal waiting
  time corresponds to $a_{2}^{\ini}\simeq a_{2}^{\hcs}$ \cite{PyT13}.
\bibitem{additional_friction} The same property holds when an
  additional friction force, coming from the thermal bath, is taken
  into account \cite{CVyG13}. Thus, the qualitative behavior discussed
  below is expected to remain valid for  that case.
\end{thebibliography}
\end{document}